\documentclass[11pt,twoside]{article} % Specifies the document style.
\usepackage{times,fancyhdr}
\usepackage{cite}
\usepackage{graphicx}

\usepackage{bm}
\usepackage{amssymb}
\usepackage{hyperref}

\def\la{\label}
\def\beq{\begin{equation}}
\def\eeq{\end{equation}}
\def\bea{\begin{eqnarray}}
\def\eea{\end{eqnarray}}
\def\p{\partial}

\date{}

\title{Coherent oscillations in superconducting cold Fermi atoms and their
applications} % Declares the document's title.
\author{Razvan Teodorescu \\ Theoretical Division and Center for Nonlinear Studies, \\ Los Alamos National Laboratory}

\begin{document} % End of preamble and beginning of text.

\pagestyle{fancy}
\fancyhead{} % clear all header fields
\fancyhead[EC]{Razvan Teodorescu}
\fancyhead[EL,OR]{\thepage}
\fancyhead[OC]{Coherent oscillations in superconducting cold Fermi atoms and their
applications}
\fancyfoot{} % clear all footer fields
\renewcommand\headrulewidth{0.5pt}
\addtolength{\headheight}{2pt} % make space for the rule

\abstract{Recent achievements in experiments with cold fermionic atoms indicate the potential
for developing novel superconducting devices which may be operated in a wide
range of regimes, at a level of precision previously not available. Unlike traditional,
solid-state superconducting devices, the cold-atom systems allow the fast switching
on of the BCS phase, and the observation of non-equilibrium, coherent oscillations of
the order parameter. The integrable and non-linear nature of the equations of motions
makes this operating regime particularly rich in potential applications, such as quantum
modulation and encoding, or nonlinear mixing of quantum coherent oscillations,
to name only two.
From a mathematical point of view, such systems can be described using
the Knizhnik-Zamolodchikov-Bernard equation, or more generally, by the matrix
Kadomtsev-Petviashvilii integrable hierarchy. This identification is particularly useful,
since it allows a direct application of the known non-linear phenomena described
by particular solutions of these equations. Other important features of this formulation,
such as the relation to the spin Calogero-Sutherland model, also have relevant
physical interpretations. In this work, a complete description of these relationships is
presented, along with their potential practical consequences.}

\maketitle % Produces the title.

%\hfill{LA-UR-07-2504}

\section{Introduction}

The operating parameters specific to experiments with cold fermionic
gases exhibiting the paired BCS state \cite{Fermi1,Fermi2,Fermi3,Fermi4,Fermi5,Fermi6,Fermi7,Fermi8}  allow the 
investigation of previously inaccessible 
 regimes, as suggested in \cite{Levitov}. It was shown 
that for such systems, the time scales of the order parameter $\tau_{\Delta}
\sim |\Delta|^{-1}$, and the quasiparticle energy relaxation time $\tau_{\epsilon}$ are both much larger than typical time for switching
on the pairing interaction $\tau_0$, essentially given by the variation of external parameters, such as detuning from the Feshbach resonance. It was 
argued that in this regime, for times $t \ll \tau_{\epsilon}$, the dynamics
of the system is given by non-linear, non-dissipative equations describing the coherent BCS fluctuations for the system out of equilibrium. In this limit, the system is integrable, and features non-perturbative behavior, such as soliton-type solutions.

In the mean-field limit, such non-trivial solutions describing the collective mode of the Anderson spins \cite{Anderson} were derived in \cite{Levitov}, for a two-level effective system. This work was generalized \cite{YAKE04} in algebro-geometrical terms. 

In \cite{Gurarie,Leggett,YTA05}, the long-time behavior of the solution has been considered, under various conditions. An issue that remains incompletely addressed  
so far is the relaxation of the nonlinear oscillatory solution induced by perturbations of the spectral curve, physically justified by coupling to the environment. Several possible kinds of perturbations may be considered, which may lead to different types of relaxation.
 
In this article, we describe the mathematical structure underlying the integrable 
quantum pairing problem, and its relations with the theory of isomonodromic deformations
of differential equations and other integrable models of strongly interacting particles. 
We also explain the classical limit of the quantum hamiltonian, and give the explicit 
construction of the solution in this case. Possible applications due to the integrability of the model are listed in the conclusion of the article.

The paper is organized as follows: in section \ref{first}, the Richardson-Gaudin model and its relation to other integrable models are reviewed. Section \ref{second} describes the classical limit and explicit multi-phase solution. In section \ref{third}, a few possible applications are briefly discussed.

\section{The Richardson-Gaudin Model} \la{first}

As argued for the first time in \cite{Levitov}, optically trapped cold Fermi gases in the BCS state are correctly described by the Richardson-Gaudin quantum pairing model. Using this model is justified by the discreteness of the energy spectrum, and by the spatially uniform profile characterizing such systems. This integrable system, introduced by Richardson and Sherman \cite{Richardson1,Richardson2,Richardson3,Richardson4,Richardson5,Richardson6} in the context of nuclear physics, has received revived interest in recent years, after being applied to metallic  superconducting grains 
\cite{Sierra}. The model is intimately related \cite{CRS97,ALO21,DES01} to a class of integrable systems 
generally referred to as Gaudin magnets \cite{Gaudin76}. These systems have been 
studied both at quantum and classical level \cite{Dukelsky,AFS,Samtleben,Hikami,Gawedzki1,Gawedzki2,Ortiz,Korotkin,Harnad,Dilorenzo}, in the elliptic case as well as trigonometric and rational degenerations, using various methods  from integrable vertex models to singular limits of Chern-Simons theory. 

In this section, we recall the main features of this model, as well as their physical interpretations.

\subsection{The quantum pairing hamiltonian}

Following \cite{Dukelsky}, we briefly review the Richardson pairing model.  
It describes a system of $n$ fermions characterized by a set of independent one-particle states of energies $\epsilon_l$, where the label $l$ takes values from a set $\Lambda$. The labels may refer, for instance, to orbital angular  momentum eigenstates. Each state $l$ has a total degeneracy $d_l$, and the states within the subspace corresponding to  $l$ are further labeled by an internal quantum number $s$. For instance, if the quantum number $l$ labels orbital momentum eigenstates, then $d_l = 2l+1$ and $s = -l, \ldots, l$. However, the internal degrees of freedom can be defined independently of $l$. We will assume
that $d_l$ is even, so for every state $(ls)$, there is another one related by time reversal symmetry $(l\bar s)$. For simplicity, we specialize to the case $d_l = 2, s= \uparrow, \downarrow $. Let $\hat c^{\dag}_{ls}$ represent the fermionic creation operator for the state $(ls)$. Using the Anderson pseudo-spin operators \cite{Anderson}  (quadratic pairing operators), satisfying the $su(2)$ algebra
\beq
[t^3_i, \, t_j^{\pm}] = \pm \delta_{ij} t_j^{\pm}, \,\,\,\, [t^+_i, \, t^-_j] = 2\delta_{ij}t^3_j,
\eeq
the Richardson pairing hamiltonian is given by
\beq \label{pairing}
H_P = 
\sum_{l \in \Lambda} 2\epsilon_l t^3_l - g\sum_{l, l'}t^+_l t^-_{l'} = 
\sum_{l \in \Lambda} 2\epsilon_l t^3_l - g{\bf{t}}^+\cdot {\bf{t}}^-, 
\eeq
where ${\bf {t}} = \sum_{l}{\bf {t}}_l$ is the total spin operator. 
It maps to the  reduced BCS model 
\beq \label{hamiltonian}
\hat H = \sum_{{\bf{p}}, \sigma}\epsilon_{\bf{p}}\hat{c}_{{\bf{p}}, \sigma}^{\dag}\hat{c}_{{\bf{p}}, \sigma}
-g\sum_{{\bf{p}}, {\bf{k}}}\hat{c}_{{\bf{p}} \, \uparrow}^{\dag}\hat{c}_{{-\bf{p}}\, \downarrow}^{\dag}
\hat{c}_{{-\bf{k}}\, \downarrow}\hat{c}_{{\bf{k}}\, \uparrow}
\eeq
by replacing the translational degrees of freedom by rotational ones, where $l  \in \Lambda = \{ 1, \ldots n \}$ ennumerates the one-particle orbital degrees of freedom, while $s = \, \uparrow, \, \downarrow$ indicates the two internal spin states per orbital ($d_l=2$). The pairing hamiltonian can be decomposed into the linear combination
\beq
H_P=2\sum_{l \in \Lambda} \epsilon_l R_l + g \left [ \left ( \sum_{l \in \Lambda} t^3_l \right )^2 - \frac{1}{4} \sum_{l \in \Lambda} (d_l^2-1) \right ].
\eeq
At a fixed value of the component $t^3$ of the total angular momentum, the last term becomes a constant and is dropped from the 
hamiltonian. The operators $R_l$ (generalized Gaudin magnets \cite{Gaudin76}) are given by
\beq \label{gaudin}
R_l = t^3_l - \frac{g}{2}\sum_{l' \ne l}\frac{{\bf {t}}_l \cdot {\bf {t}}_{l'}}{\epsilon_l - \epsilon_{l'}}.
\eeq
These operators solve the Richardson pairing hamiltonian because \cite{CRS97} they are independent, commute with
each other, and span all the degrees of freedom of the system. Richardson showed \cite{Richardson1,Richardson2} that the exact $N-$pair wavefunction 
of his hamiltonian is given by application of operators 
%\beq
$
b^{\dag}_{k} = \sum_{l}\frac{t_l^{\dag}}{2\epsilon_l - e_k}
$
%\eeq 
to vacuum (zero pairs state). The unnormalized $N-$pair wavefunction reads
%\beq
$
\Psi_R(\epsilon_i) = \prod_{k=1}^Nb_k^{\dag}|0\rangle.
$
%\eeq 
The eigenvalues $e_k$ satisfy the self-consistent algebraic equations 
\beq \la{ec}
\frac{1}{g} = \sum_{p \ne k}\frac{2}{e_k - e_p} +\sum_l \frac{1}{2\epsilon_l-e_k},
\eeq which can be given a 2D electrostatic interpretation \cite{Dukelsky}
with energy
\begin{eqnarray} \label{electrostatic}
U(\epsilon_l, e_k) = \frac{2}{g} \left [ \sum_{k=1}^N 
\mathcal{R}e (e_k) -
\sum_{l=1}^n \mathcal{R}e (\epsilon_l) \right ] + \\
2\sum_{l=1}^n\sum_{k=1}^N \log|e_k -2\epsilon_l| - 
4\sum_{k < p}\log|e_k-e_p|- 
\sum_{i<j}\log|2\epsilon_i-2\epsilon_j|
\end{eqnarray}
Equations (\ref{ec}) appear as equilibrium conditions for a set of charges of strength $q=2$ placed at points $e_k$, in the presence of fixed charges of strength $q=-1$ at points 
$2\epsilon_l$, and uniform electric field of strength $\frac{1}{g}$, pointing along the real axis. This interpretation proves to be very useful for the conformal field theory (CFT) description of the Richardson problem. The electrostatic energy 
(\ref{electrostatic}) is minimized for values $\{ e_k\}$ corresponding to pair energies. In (\ref{electrostatic}), $n, N$ represent the number of single-particle levels and the number of pair energies, respectively. 
For large interaction constant $g$, the equilibrium positions $\{ e_k \}$ form a set of complex conjugated pairs defining a curve $\gamma$ in the complex plane of energies. We note that the eigenvalues $r_i$ of Gaudin hamiltonia are proportional in this language to the values of the electric field at positions $2\epsilon_i$, $2r_i = g\frac{\p U}{\p \epsilon_i}. $

For a set of single-particle energies $\{\epsilon_i \}$, the BCS ground state is obtained by minimizing the electrostatic energy (\ref{electrostatic}) with respect to positions of
the free charges at $\{ e_k \}$. Once found,
they also determine exactly the values of the electric field at positions
$\{ 2\epsilon_i \}$ on the real axis, which are proportional to 
the ground-state eigenvalues  $\{ r^{GS}_i\}$. For any other values of $\{ r_i \}$, the electrostatic energy 
(\ref{electrostatic}) is not minimized. This indicates
that for arbitrary values $r_i \ne r_i^{GS}$, the system is not in equilibrium.

\subsection{Richardson-Gaudin model as limit of KZB equations}

In this section we recall the relation between the Richardson-Gaudin model and
singular $SU(2)$ Chern-Simons and WZW models \cite{Gawedzki1,Gawedzki2}. These relations stem from the study of quantum states of the Chern-Simons theory with gauge group $SU(2)$ on the manifold $\mathbb{T}^2 \times \mathbb{R}$ and in the presence of Wilson lines $\{ z_i\} \times \mathbb{R}$, $i = 1, \ldots, n$ \cite{Gawedzki1}, corresponding to the one-particle spectrum of the Richardson hamiltonian,
$z_i = 2\epsilon_i$. The torus $\mathbb{T}^2$ has modular parameter $\tau = \tau_1 + i\tau_2, \,\, \tau_{1,2} \in \mathbb{R}$. Thus, $\mathbb{T}^2 = \mathbb{C}/\mathbb{Z} + \tau\mathbb{Z}$. The quantum states of this theory are known to satisfy the Knizhnik-Zamolodchikov-Bernard (KZB) equations 
\beq \label{kzb}
\nabla \Psi_{CS} = 0, \quad \nabla = (\nabla_{\tau}, \nabla_{z_i}), 
\eeq
where $\nabla$ is the flat KZB connection,
\beq \label{kzbcon}
\nabla_{\tau} = \kappa \p_{\tau} + H_0(\tau, z_i), \quad 
\nabla_{z_i} = \kappa \p_{z_i} + H_i(\tau, z_i), 
\eeq
and $H_0, H_i$ are elliptic versions of the Gaudin hamiltonia. The parameter 
$\kappa$ is related to the level of representation of $SU(2)$. By a limiting procedure, the system degenerates into the Richardson-Gaudin problem: by taking the limit 
$\tau \to i\infty$, the torus degenerates into a cylinder, and the elliptic hamiltonia $H_i, H_0$ take a trigonometric limit. Upon rescaling the cylinder such that the set $\{z_i\}$ is collapsed into the origin, the cylinder becomes the complex plane with punctures at $\{z_i\}$, and the rescaled operators become the rational Gaudin magnets (\ref{gaudin}). 

\subsection{Singular representations of Wess-Zumino-Witten theory} 

Given the well-known relation between $2+1$ Chern-Simons theory with $SU(2)$ gauge group and WZW theory, it is not surprising that a non-Abelian CFT representation for Richardson-Gaudin models is also available. In fact, the relationship between these theories is already transparent by identifying the electrostatic picture of R-G with a Coulomb gas representation of WZW. 

In the CFT approach \cite{Sierra}, a primary field of spin $1/2$ of the $SU(2)$ representation of WZW is associated to every energy level of the R-G problem. 
The background charge $\alpha_0$ and the level of Kac-Moody algebra $SU(2)_k$
are related by $k = \frac{1}{2\alpha_0^2} - 2$. It turns out that the R-G model is retrieved in the singular limit $\alpha_0 \to \infty$, corresponding to $k=-2$. In this limit, the representations are given by $\it{perturbed}$ WZW blocks (PWZW), 
\beq \label{pwzw}
\Psi_{P} (z_i) = \oint_{\mathcal{C}_1}d e_1 \oint_{\mathcal{C}_N} d e_N
e^{-\alpha_0^2 U(z_i,e_k)}\Psi_R(z_i, e_k). 
\eeq  
In (\ref{pwzw}), $e_k$ are the eigenvalues of the Richarsdon hamiltonian with 
$N$ fermion pairs, $\alpha_0$ is the background charge in the Coulomb gas representation of the WZW model associated with the $2+1$ Chern-Simons theory described above, and $U(z_i,e_k)$ is the holomorphic piece of the 2D electrostatic potential of the Coulomb gas representation (\ref{electrostatic}). The Richardson wave function $\Psi_R(z_i, e_k)$ is obtained 
by applying Bose fields corresponding to positions $z_i, e_k$,
\beq
\Psi_R(z_i, e_k) = \langle \prod_{i=1}^n\gamma(z_i)\prod_{k=1}^N\beta(e_k)\rangle,
\eeq 
and with correlator $\langle\beta(z)\gamma(z')\rangle=\frac{1}{z-z'}$. 

The singular limit described above in the language of the Chern-Simons theory corresponds to the limit $\alpha_0 \to \infty$ in the CFT approach. The perturbed WZW blocks are dominated by the saddle-point configurations, given by the eigenvalue equations 
$\p_{e_k}U = 0$, and become proportional to the Richardson wavefunction. In 
the singular limit, the KZB equations takes the form
\beq \label{pkz}
\left (\frac{1}{2\alpha_0^2} \p_{z_i} - R_i \right ) \Psi_{PWZW} = 0,
\eeq
which shows that the Richardson wavefunction is an eigenstate of the Gaudin hamiltonia, with eigenvalues given by the solutions to the electrostatics problem (\ref{electrostatic}). 

Equation (\ref{pkz}) is the analog of the singular limit of the KZB equation (\ref{kzb}), in the limit $\kappa \to 0$, corresponding to $\alpha_0 \to \infty$. Likewise, there is a dirrect correspondence between the Chern-Simons states and the perturbed WZW blocks. We will see in the following sections that these equations survive in the semiclassical limit and that they allow the identification of the tau function for the corresponding classical integrable system.

\subsection{Relation with the KP hierarchy and other integrable models} 

The zero curvature conditions for the KZB connection can be interpreted as a truncation 
of the general Kadomtsev-Petviashvilii (KP) integrable hierarchy, at a lelvel set by the 
number of eigenvalues in the single-particle spectrum. This is also illustrated by 
the fact that, as shown in the preceding paragraphs, in the genus one case, the Richardson-Gaudin problem is equivalent with the elliptic Calogero model. This model is known to be embedded into the KP hierarchy, and admits simple-pole solutions built from the solution of KP. In fact, all the relationships described so far are consequences of an even more general formulation of the Richardson-Gaudin problem as a hierarchy of isomonodromic deformations with respect to changing the moduli of its spectral curve. 

These relationships allow to classify the solutions to the quantum Richardson-Gaudin model using degenerate limits of the solutions for the elliptic Calogero problem. Such solutions are known (for instance, the simple-pole solutions mentioned above) and would therefore 
allow (in principle) constructing exact solutions for given initial conditions. Even more importantly, there is a large body of knowledge concerning the classes of solutions supported by the KP hierarchy (periodic, algebraic-geometric or rational), which translates into distinct "phases" (albeit we are dealing with non-equilibrium, coherent oscillations) supported by the quantum pairing model, depending on the initial conditions.

\section{The mean-field limit of Richardson-Gaudin models} \la{second}

Throughout this section, we consider the classical limit of the quantum pairing 
model. The main reason for this approximation is the fact that explicit, generic
solutions can be obtained in this case. 

\subsection{General description of the classical model} 
In  the mean-field limit,
the spin operators ${\bf {t}}_l$ are replaced by their quantum mechanical averages. Written in terms of the classical vectors ${\bf {S}}_l = 2 \langle {\bf {t}}_l \rangle$, the semiclassical approximation for the pairing 
hamiltonian becomes
\beq \label{meanfield}
H_{MF} = \sum_{l \in \Lambda}\epsilon_l S^3_l - \frac{g}{4}|J^-|^2,
\eeq 
where ${\bf {J}} = \sum_{l \in \Lambda} {\bf {S}}_l$ and the BCS gap function is given by $\Delta = gJ^-/2$. Replacing commutators  by canonical Poisson brackets, 
\beq \label{poisson}
\{ S^{\alpha}_i, \, S^{\beta}_j \} = 2\epsilon^{\alpha \beta \gamma}S^{\gamma}_i \delta_{ij},
\eeq
variables $S^{\alpha}_i$ become smooth functions of time. In this limit, 
the problem can analyzed with tools of classical integrable systems, and the 
solution is known to be exact as $n \to \infty$. 

The Poisson brakets (\ref{poisson}) and hamiltonian 
(\ref{meanfield}) lead to the equations of motion
\beq \label{bloch}
\dot{\vec{{S}}}_i = 2(-\vec{\Delta} + \epsilon_i \hat{z}) \times \vec{S}_i,
\eeq
where $2\vec{\Delta} = (gJ_x, gJ_y, 0)$ and $\vec{J}$ is the total spin. The 
semiclassical limit of Gaudin hamiltonia are independent constants of motion,
\beq
r_i = \frac{1}{2}\left [S_i^z - \frac{g}{2}\sum_{j \ne i}
\frac{\vec{S}_i \cdot \vec{S}_j}{\epsilon_i -\epsilon_j} \right ],
\quad
%\{r_i, \, r_j \} =0, \quad
%\eeq
%\beq
\dot r_i 
%= \{ 2\sum_j \epsilon_j r_j, \, r_i \} 
= 0.
\eeq
Equations (\ref{bloch}) describe a set of strongly interacting spins and have
generic non-linear oscillatory solutions. The exact solution may be obtained through the Abel-Jacobi inverse map, explained in the next section.

\subsection{Isospectral case and the Abel-Jacobi inversion problem}

\begin{figure} \begin{center}
 \includegraphics*[width=8cm]     {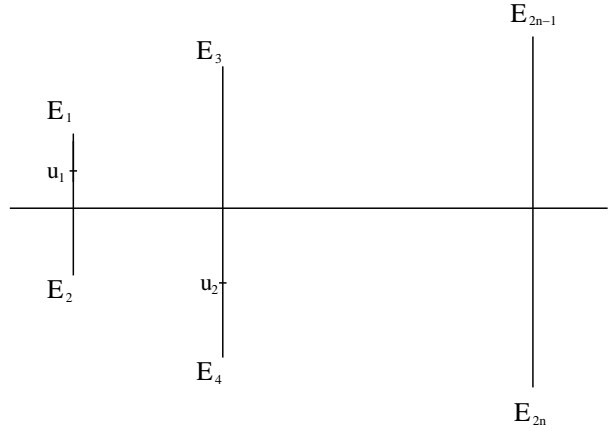}
\caption{\label{oscilat}
Schematic representation of the Liouville torus for the integrable
system (\ref{bloch}).
}
\end{center} \end{figure}

In \cite{Kuznetzov,YAKE04}, the system (\ref{bloch}) with fixed spectral curve was solved through inverse Abel-Jacobi mapping, by using Sklyanin separation of variables techniques \cite{SOV,Smirnov}. 
Interesting connections to generalized Neumann systems and Hitchin systems were discovered in \cite{Hikami}.
The solution starts from the Lax operator 
\beq \label{lax} 
\mathcal{L}(\lambda) = \frac{2}{g}\sigma_3 +\sum_{i=0}^n \frac{\vec S_i \cdot\vec \sigma}{\lambda - \epsilon_i} = 
\left [
\begin{array}{cr}
a(\lambda) & \,\, \, b(\lambda) \\
c(\lambda) & -a(\lambda)
\end{array}
\right ],
\eeq
where $\sigma_\alpha, \, \alpha = 1,2,3$ are the Pauli matrices, and $\lambda$ is an additional complex variable, the spectral 
parameter. Let $u_k, \, k=1, \ldots, n-1$ be the roots of the coefficient $c(\lambda)$. 
Poisson brackets for variables $S_i^{\alpha}$ read 
\beq
\{S_j^\alpha , \, S_k^{\beta} \}=2\epsilon_{\alpha \beta \gamma}S_k^\gamma \delta_{jk}
\eeq
The Lax operator (\ref{lax}) defines 
a Riemann surface (the spectral curve) $\Gamma (y, \lambda)$ of genus 
$g = n-1$, through 
\beq \label{curve}
y^2 = Q(\lambda) = \det \mathcal{L}(\lambda)\left [ g\frac{P(\lambda)}{2} \right ]^2 , 
\eeq
where $P(\lambda) = \prod_{i=1}^n (\lambda -  \epsilon_i).$

The equations of motion for the hamiltonian (\ref{meanfield}) become
\beq \label{dubrovin1}
\dot  u_i = \frac{2i y( u_i)}{\prod_{j \ne i}( u_i - u_j)},
\quad
i\dot J^- = J^{-} \left [ gJ^3 + 2\sum_{k=1}^{n} \epsilon_k - u \right ].
\eeq
In (\ref{dubrovin1}), 
$u = -2\sum_{i=1}^{n-1}  u_i, \quad b(u_i) = 0$. 

From the equations of motion, it is clear that knowledge of the initial amplitude of $J^-$ and of the roots $ \{  u_i \}$
is enough to specify the $n$ unit vectors $\{ {\bf {S}}_i \},$ for a given set of constants of motion $\{ R_l \}$ given by the classical limit of Gaudin hamiltonia. The Dubrovin equations (\ref{dubrovin1}) are 
solved by the inverse of the Abel-Jacobi map, as we explain in the following. We begin by noting that the polynomial 
$Q(\lambda)$ has degree $2n$, and is positively defined on the real $\lambda$ axis. Therefore, the curve $\Gamma (y, \lambda)$
has $n$ cuts between the pairs of complex roots $[E_{2i-1}, \, E_{2i}], i = 1, 2, \ldots, n$, perpendicular to the real $\lambda$
axis. The points $u_i$ belong to $n-1$ of these cuts, $u_i \in [E_{2i-1}, \, E_{2i}], i =1, \ldots, n-1$. These 
$g = n-1$ cuts allow to define a canonical homology basis of $\Gamma$, consisting of cycles $\{\alpha_i, \beta_i \}, i = 1, \ldots, g$. With respect to these cycles, a basis of normalized holomorphic differentials $\{ \omega_i \}$ can be defined, through
\beq
\mu_i = \lambda^{g-i}\frac{d\lambda}{y}, \,\,\, M_{ij} = \int_{\alpha_j} \mu_i, \,\,\, {\bm {\omega}} = M^{-1}{\bm{\mu}}. 
\eeq 
The period matrix $B_{ij} = \int_{\beta_j} \omega_i$
is symmetric and has positively defined imaginary part. The Riemann $\theta$ function is defined with the help of the period matrix as
\beq
\theta({\bm{z}} | B) = \sum_{{\bm{n}} \in {\bm{z}}^g} e^{2\pi i ({\bm {n}}^t {\bm{z}} + \frac{1}{2} {\bm {n}}^t B {\bm {n}} )}.
\eeq
The $g$ vectors $\bm{B}_k$ consisting of columns of $B$ and the basic
vectors $\bm{e}_k$  define a lattice in ${\mathbb{C}}^g$. The $Jacobian$ variety of the curve $\Gamma$, is then the $g-$dimensional torus defined as the quotien 
$J(\Gamma)  = \mathbb{C}^g /(\mathbb{Z}^g + B\mathbb{Z}^g).$
The Abel-Jacobi map associates to any point $P$ on $\Gamma$, a point ($g-$ dimensional complex vector) on the Jacobian variety, through
${\bm {A}} (P) = \int_{\infty}^P {\bm {\omega}}.$
Considering now a $g-$dimensional complex vector of points $\{P_k \}, 
k = 1, \ldots, g$ on $\Gamma$, defined up to a permutation, we can associate to it the point on the Jacobian
\beq  \label{a}
{\bm{z}} = {\bm{a}}({\bm {P}}) = \sum_{k=1}^g {\bm {A}} (P_k) + {\bm {K}},
\eeq
where ${\bm {K}}$ is the Riemann characteristic vector for $\Gamma$. 

The map (\ref{a}) suggests that we now have a 
way to describe the dynamics on $\Gamma$ by following the image point on the Jacobian. 
Given a point on the $g-$dimensional Jacobian  ${\bm{z}} = (\zeta_1, \ldots , \zeta_{n-1})$, we can find an unique set of points $\{ \lambda_k\}, k = 1, \ldots, g$ on $\Gamma$, such that ${\bm{z}} = {\bm {a}}(
{\bm {\lambda}}),$ and $\theta ({\bm{a}(\bm{P}) - \bm{z}} | B) = 0$. The system evolves in time according to the point ${\bm{z}}(t)$ 
\beq \label{solution1}
\zeta_k = ic_k, \,\, 1 \le i \le g-1, \,\, \zeta_{n-1} = i(c_{n-1} + t),
\eeq
where $\{ c_k\}$ is a set of initial conditions, such that  $
{\bm{z}}_0 = {\bm{z}}(t=0) = {\bm {a}}({\bm{c}})$, and ${\bm{c}}$ is the set of initial
conditions for positions of ${\bm {\lambda}}$ on $\Gamma$. Together with the initial condition which determines the
initial amplitude of $J^-$, this set will determine entirely the evolution of the functions $ u_i(t), J^-(t)$.

It is perhaps useful to give a graphical description (Figure~\ref{oscilat}) of the general solution described above. For a given set of initial conditions for the spins 
$\{ \vec{S}_i \}$, $i.e.$ also of the constants of motion $\{ r_i \}$, a
polynomial $Q(u)$ of degree $2n$ and with $n$ pairs of complex conjugated
roots $E_{2k+2}=\overline{E}_{2k+1}, \, k=0, \ldots, n-1$, is constructed. 
A schematic representation of these roots is given in Figure~\ref{oscilat}. 
Between each pair of roots, we place a simple cut 
$\mathcal{C}_k = [E_{2k+1}, \, E_{2k+2}]$ 
on the complex plane of energies. The surface thus obtained is a representation 
of a torus of smooth genus $g=n-1$.  

Variables $u_k$ are introduced for $n-1$ of these cuts, with respect to which the equations of motion separate. The variables $u_k$ evolve in time in a complicated fashion, solving a system of nonlinear coupled differential equations (\ref{dubrovin1}). Up to a constant, the time dependence of the gap parameter amplitude is given by 
\beq
\log |\Delta(t)| = \mathcal{I}m \, \int u(t) dt, \quad u(t) = 
\sum_k u_k (t). 
\eeq
The widths of the cuts $\mathcal{C}_k$ and the 
periods of the non-linear oscillators $u_k$ are determined by the values of
constants of motion $\{ r_i \}$. For the particular choice $r_i =r_i^{GS}$, all
the cuts $\mathcal{C}_k, \, k=1, \ldots , n-1$ vanish, and the width of the remaining cut $\mathcal{C}_n$ equals the equilibrium value of the gap function:
$|E^{GS}_{2n-1}-E^{GS}_{2n}| = 2 |\Delta|^{GS}$. In that case, the oscillators
$u_k=E_{2k-1}=E_{2k}$ are at rest, and the only time dependence left in the system is the uniform precession of the parallel planar spins $S^{-}_i$, with 
frequency $\omega = 2\sum_{k=1}^n\epsilon_k - 2\sum_{p=1}^{n-1}E_{2p-1}-\sum_{i=1}^nS^z_i$. In the case of particle-hole symmetry, $\omega$ vanishes as well. 

\section{Conclusion} \la{third}

In the present work, the regime of coherent, non-equilibrium quantum 
oscillations characterizing the onset of the BCS state in cold Fermi gases 
was investigated from a mathematical point of view. The integrability of
the model and main features of the space of solutions were described, both 
at the quantum and classical levels. 

The most direct application of these coherent quantum oscillations would be in
encoding/decoding of information. Although simple in principle, this would in 
fact require very accurate control over the initial conditions at the onset of
 the pairing regime, which is unfeasible. However, given the stability of the coherent oscillatory regime, it is clearly easier to first implement the quantum 
modulation of an arbitrary oscillatory solution, by slowly perturbing the 
single-particle spectrum. This kind of perturbation is in fact within the same class as the original isomonodromic deformations, and hence does not destroy the integrability of the model. 

Another possible application makes use of the non-linear nature of the solution 
of the Richardson-Gaudin model. Considering only the simplest classical solutions given by the Abel-Jacobi construction, the elliptical solutions, it is possible to devise a 
non-linear 'mixer' by using the fact that the dynamics is linear on the associated torus (Jacobian). Then, linear operations with respect to the dynamics on the Jacobian would translate into strongly non-linear (but still invertible) effects observed at the level of quantum oscillations.

\subsection*{Acknowledgments}

This work was carried out under the auspices of the National Nuclear Security 
Administration of the U.S. Department of Energy at Los Alamos National 
Laboratory under Contract No. DE-AC52-06NA25396.

\end{document}